\documentclass{eas}
\usepackage{graphicx}
\begin{document}

\title{Calibrating the Cepheid Period-Luminosity relation with the VLTI} 
\runningtitle{VLTI and Cepheid P-L relation}
\author{Massimo Marengo, Margarita Karovska, Dimitar Sasselov,
Costas Papaliolios}\address{Harvard-Smithsonian Center for
Astrophysics, 60 Garden St., Cambridge, MA 02138, USA} 
\author{J.T. Armstrong}\address{Naval Research Laboratory, Code 7210,
Washington DC 20375, USA}
\author{Tyler E. Nordgren}\address{University of Redlands, 1200 E
Colton Ave, Redlands, CA, USA}
\begin{abstract}
The VLTI is the ideal instrument for measuring the distances of
nearby Cepheids with the Baade-Wesselink method, allowing an accurate
recalibration of the Cepheid Period-Luminosity relation. The high
accuracy required by such measurement, however, can only be reached
taking into account the effects of limb darkening, and its dependence
on the Cepheid pulsations. We present here our new method to compute
phase- and wavelength-dependent limb darkening profiles, based on
hydrodynamic simulation of Classical Cepheid atmospheres. 
\end{abstract}
\maketitle

\section{Introduction}

Since the discovery of their Period-Luminosity (P-L) relation, at the
beginning of the century (Leavitt \cite{leavitt1906}), Classical
Cepheids have played a central role as primary distance indicators. 
The reliability of this relation, however, depends on the accuracy
of its calibration. This is done by using a handful of nearby
Cepheids, whose distances can be independently determined. Due to the
low-density distribution of the Cepheids in our galaxy, however, this
is a difficult task to accomplish. Large errors still affect direct
parallax measurements done with the Hipparcos satellite (see
e.g. Madore \& Freedman \cite{madore1998}), and even future
space-based surveys appear to suffer from the same limitations. 

Recent progresses in ground-based, long baseline interferometry,
however, have finally opened the possibility of direct measurements
of nearby Cepheids distances by using the so-called \emph{geometric}
Baade-Wesselink (BW) method (Baade 1926, Wesselink 1946). A detailed
description of this method and its implications for the accuracy of
the zero-point calibration of the Cepheid P-L relation is given in
Sasselov \& Karovska \cite{sasselov1994a} (SK94 hereafter). 

The BW method relies on the accurate measurement of the Cepheid
diameter as the star pulsates. As such, it can only be carried on with
high angular resolution interferometric measurements. The VLTI will be
the ideal instrument for such measurements, given the available
long baselines, and the large apertures allowing to reach fainter
sources. However, to convert the measured VLTI visibilities into
accurate diameters, it is critical to use appropriate limb darkening
models for different pulsational phases.

\begin{figure}
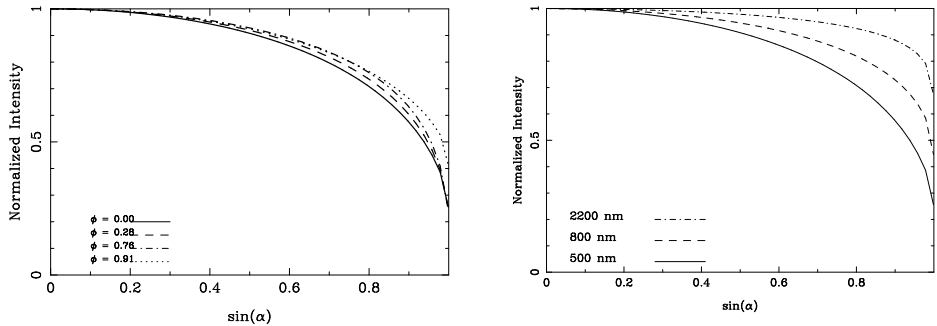

\begin{center}
\includegraphics[height=0.47\textwidth,angle=-90]{fig1aa.ps}
\qquad
\includegraphics[height=0.45\textwidth,angle=-90]{fig1b.ps}
\end{center}
\caption{Limb darkened profiles at different pulsational
phases (\emph{left}) and at different wavelengths (\emph{right})
for our $\zeta$ Gem model. Left panel profiles are computed for
$\lambda$ = 500 nm; right panel profiles for minimum radius
($\phi$ = 0). Adapted from MM02.}\label{fig-1}
\end{figure}

\section{Limb darkening models for pulsating Cepheids}

As described in SK94, the feasibility of the geometric BW method
depends on the availability of accurate hydrodynamic models for the
Cepheid atmospheres, which are required to compute reliable
time-dependent limb darkening profiles. In Marengo et
al. \cite{marengo2002} (MM02 hereafter)
we describe a new approach to compute 
detailed wavelength- and phase-dependent limb intensity profiles for
pulsating Cepheids. Our profiles are based on time-dependent, non-LTE
hydrodynamic computations which, starting from a model of the Cepheid
pulsation, reproduce the dynamic structure of the stellar atmosphere
as it pulsates (see Sasselov \& Lester \cite{sasselov1994b}). From
these model atmospheres, we obtain the intensity emerging from
the stellar photosphere (and thus the limb darkening), computed for
each pulsational phase relevant to the interferometric
observations. This procedure can be applied to all Cepheids for which
a reliable pulsational model is available.

Figure~\ref{fig-1} shows the effects of non-LTE hydrodynamics in the
limb darkening of the Classical Cepheid $\zeta$~Gem. Limb darkening
appears to change with the pulsational phase and wavelength. The
change is not only in the amount of limb darkening, but also in the
shape of the limb profile. This shows that a single parametrization of
the limb profile is not adequate for all pulsational phases. Detailed
time-dependent modeling is thus required for realistic limb darkening.

Our simulations show that the corrections introduced by the
hydrodynamic effects and the wavelength dependence of the limb
darkening in the diameter measurement, are of the order of a few
percent at optical wavelengths. In the near-IR, which is the VLTI
primary domain, the corrections are of the order of $\sim$2\%,
and are still sensitive to the effects of the pulsations.

\section{Geometric BW method and the VLTI}

The high accuracy of the VLTI visibilities,
and the large spectral dispersion of the AMBER camera, will finally
provide the high quality data necessary to apply the BW method. At the
same time, VLTI  will allow the first direct testing of the
atmospheric model themselves, improving our knowledge of the stellar
atmosphere dynamics. To reach the full potential
of the VLTI, it is however necessary to take into account the
corrections induced by limb darkening and its dependence from the
pulsational hydrodynamics. 

As the only long baseline interferometer below the equator, the VLTI
offers a unique opportunity to apply the geometric BW method to
southern sources. VLTI provides both the sensitivity and the angular
resolution necessary to detect the radial variations of a sample of
Classical Cepheids, to be used to calibrate the zero point of the P-L
distance relation. 

The sensitivity required is of the order of mag 2--4 in the K band,
for stars which are of mag 5--7 in the optical. Such sensitivity can
be easily achieved at the VLTI by the near-IR camera AMBER, using a
setup with three Auxiliary Telescopes, and large spectral resolution
to isolate individual spectral lines. The typical expected diameters
are of the order of $\sim 1$ mas, with a variation due to the stellar
pulsation of about 10\% of the diameter. This angular resolution can
be reached with the VLTI by using baselines of $\sim$100m, or longer.

To-date, the angular diameter variations of Classical Cepheids have
been measured only for two sources ($\zeta$ Gem and $\eta$ Aql, see
Lane et al. \cite{lane2000}; Lane et al. \cite{lane2002}). The VLTI
will give a substantial contribution in extending this sample,
providing the necessary statistical significance for the recalibration
of the P-L relation zero point with the BW method. 

\smallskip
{\footnotesize M.K. is member of the Chandra Science Center, which is
operated under contract NAS8-39073 and is partially supported by NASA}


\end{document}